\pdfoutput=1

\documentclass[11pt]{article}

\usepackage{acl}
\usepackage{multirow} 
\usepackage{cite}
\usepackage{tikz}
\usepackage{listings}
\usepackage{mdframed}
\lstnewenvironment{code}{\lstset{language=Python}}{}
\usetikzlibrary{positioning}
\usepackage{algpseudocode}
\usepackage{amsmath,amssymb,amsfonts}
\usepackage{algorithm}
\usepackage[algo2e]{algorithm2e}
\usepackage{graphicx}
\usepackage{float}
\newlength{\leftbarwidth}
\usepackage{amsmath,amssymb,amsfonts}
\usepackage{graphicx}
\usepackage{balance}
\usepackage{mdframed}

\definecolor{light-gray}{gray}{0.80}
\usepackage{textcomp}
\usepackage{algorithm,algpseudocode}
\renewenvironment{quote}%
{\begin{leftbar}\noindent\hspace{-\leftbarwidth}\xspace}%
{\end{leftbar}}%
\usepackage{amsmath}

\usepackage{hyperref}
\usepackage{cleveref}
\usepackage{xcolor}
\usepackage{booktabs}

\usepackage{verbatim}
\usepackage{xspace}
\usepackage{subfig}
\usepackage{blindtext}

\newlength{\leftbarsep}
\colorlet{leftbarcolor}{black!120}

\usepackage{framed}
\renewenvironment{leftbar}{%
    \MakeFramed {\advance \hsize -\width \FrameRestore }%
}{%
    \endMakeFramed
}
\usepackage{listings}
\usepackage{textcomp}
\usepackage{xcolor}
\def\BibTeX{{\rm B\kern-.05em{\sc i\kern-.025em b}\kern-.08em
    T\kern-.1667em\lower.7ex\hbox{E}\kern-.125emX}}
    
\definecolor{codegreen}{rgb}{0,0.6,0}
\definecolor{codegray}{rgb}{0.5,0.5,0.5}
\definecolor{codepurple}{rgb}{0.58,0,0.82}
\definecolor{keywordsColor}{rgb}{0.000000, 0.000000, 0.635294}
\definecolor{backcolour}{rgb}{255,255,255}
\lstdefinestyle{mystyle}{
  backgroundcolor=\color{backcolour}, commentstyle=\color{codegreen},
  keywordstyle=\color{magenta},
  numberstyle=\tiny\color{codegray},
  stringstyle=\color{codepurple},
  basicstyle=\ttfamily\footnotesize,
  breakatwhitespace=false,         
  breaklines=true,                 
  captionpos=b,                    
  keepspaces=true,                 
  numbers=left,                    
  numbersep=5pt,                  
  showspaces=false,                
  showstringspaces=false,
  showtabs=false,                  
  tabsize=2,
  float=tp,
  floatplacement=tbp,
  xleftmargin=1.5em,
  belowskip=-8pt
}
\usepackage{times}
\usepackage{latexsym}

\usepackage[T1]{fontenc}

\usepackage[utf8]{inputenc}

\usepackage{microtype}

\title{Language Models are Better Bug Detector Through Code-Pair Classification}

\author{Kamel Alrashedy \\
  Georgia Institute of Technology \\
  School of Computer Science \\
  Atlanta, GA, USA \\
  \texttt{kalrashedy3@gatech.edu} \\\And
  Ahmed Binjahlan \\
  Georgia Institute of Technology \\
  School of Electrical and Computer Engineering \\
  Atlanta, GA, USA \\
  \texttt{iihmto@gatech.edu} \\}

\begin{document}
\maketitle
\begin{abstract}
Large language models (LLMs) such as GPT-3.5 and CodeLlama are powerful models for code generation and understanding. Fine-tuning these models comes with a high computational cost and requires a large labeled dataset. Alternatively, in-context learning techniques allow models to learn downstream tasks with only a few examples. Recently, researchers have shown how in-context learning performs well in bug detection and repair. In this paper, we propose code-pair classification task in which both the buggy and non-buggy versions are given to the model, and the model identifies the buggy ones. We evaluate our task in real-world dataset of bug detection and two most powerful LLMs. Our experiments indicate that an LLM can often pick the buggy from the non-buggy version of the code, and the code-pair classification task is much easier compared to be given a snippet and deciding if and where a bug exists. Code and data are attached with the submission. 
Code and data are available at Github\footnote{\url{https://github.com/Kamel773/code_pair_classification}}.

\end{abstract}

\section{Introduction}

Large language models (LLMs) like GPT-3.5 \citep{GPT3} and CodeLlama \citep{Code_llama} have shown impressive capabilities in a variety of source code tasks, including code generation, bug repair, and defect prediction \citep{alrashedy}. These models have billions of parameters, which makes it difficult to fine-tune them for downstream tasks due to limited resources and the requirement for a large labeled dataset. Gathering real-world data is costly and requires human effort. However, in-context learning requires a few examples from labeled dataset where these model learn the new task without update the parameters. Recently, in-context learning has demonstrated strong performance in software engineering tasks, achieving better results in some tasks than traditional fine-tuning techniques.

\begin{figure}[h!]
\centering
 {\includegraphics[width=1.1\linewidth]{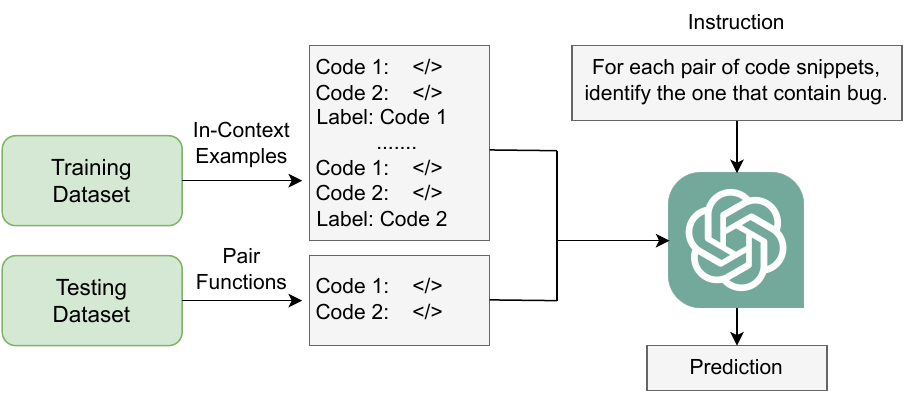}}
  \caption{Code-pair classification is an in-context learning approach in which the model receives a pair of functions and identifies the buggy one.}
  \label{fig:venn}
\vspace{-8pt}
\end{figure}

Large language models have demonstrated their capacity to generate code. Additionally, they can debug this generated code without human feedback or the use of external tools. In the real world, developers leverage tools like Co-pilot and GPT to assist in code generation. However, the produced code can occasionally be inaccurate or contain bugs, requiring human intervention for corrections. While these models can generate code, it may still contain bugs. They enhance developer productivity, handling approximately 55\% of the tasks. Nonetheless, developers must still verify the accuracy and quality of the generated code. Even though the models can debug, fix, and repair flawed code, they are not yet perfect. Developers allocate about 25–50\% of their time to debugging and testing.

The application of LLMs in binary classification tasks for bug detection has been extensively studied. Fine-tuning large language models such as CodeBERT \citep{Codebert}, CodeT5 \citep{codet5}, and PLBART \citep{PLBART} on synthetic or weakly labeled data has yielded impressive results on synthetic testing datasets. However, their performance significantly drops when applied to real-world data \citep{Saikat}. This is because real-world bugs are much more complex. For instance, in Code Snippet \ref{real-world_code}, the developers makes mistakes in calculating the denominator of the new value. Determining whether this code snippet contains a bug or not is a very challenge, even for human intelligence.

Although numerous prior studies have demonstrated progress in addressing this issue, the performance remains unsatisfactory for real-world applications. In this paper, we introduce a new task: code-pair classification. This involves providing the model with two code snippets—one containing a bug and the other the fixed version. The model's task is to identify the snippet that contains the bug.


\section{Related Work}

\textbf{LLMs for bug detection:} Applying LLM-based defect detection is an active research area in the artificial intelligence and software engineering communities \citep{great, PLUR}. \citep{self-debug} proposed self-debugging techniques where the model generates code and then debugs the generated code by itself without human feedback. The model's ability to identify and fix bugs without human intervention enhances the concept of rubber duck debugging. PLBART is a bidirectional and auto-regressive model that was pre-trained on both natural language and source code \citep{ahmad-etal-2021-unified}. This model follows the same architecture as BART, which is a sequence-to-sequence Transformer \citep{Vaswani_17}. The model was evaluated on vulnerability detection clone detection. \citep{Michael} proposed VulRepair to automatically detect and repair vulnerabilities using the T5 architecture \citep{2020t5}.

\textbf{In-context learning:} The \citep{GPT3} Introduced the concept of in-context learning, where large language models learn new tasks without updating the model's parameters. This approach has been successfully applied in many applications, such as code generation \citep{Pal} code optimization \citep{Self-refine} and comment generation \citep{Multi-Intent}. Using the concept of self-consistency in defect repair demonstrates a better improvement than the Chain of Thought (COT) approach, where the author in \citep{Self-Consistency} included commit-log messages in a few-shot setting. In \citep{zhou23docprompting}, the authors introduced DocPrompting, a novel approach that prompts the Language model using relevant documentation, enhancing to improve the accuracy of code generation. The LLM of code shows improvement in code edits and refactoring. In \citep{madaan2023learning}, the authors introduce the Performance-Improving Edits (PIE) dataset tailored for code optimization. Demonstrating a few examples of slower and faster versions of code using in-context learning, the results indicate that the LLM successfully speeds up the program. \citep{chen2022program} proposes a "Program of Thoughts" (PoT) prompt where the model generates text and code to solve complex numerical reasoning tasks.

\section{Experimental Setup}
In this section, we describe the dataset used to evaluate our approach and the chosen pretrained language models.
\subsection{Real-world dataset}
The PyPIBugs, proposed by \citep{BUGLAB}, is the largest real-world dataset for bug detection. It contains both the buggy code and its fixed version of functions from real-world applications. The authors did not release their dataset due to licensing limitations, but they provided supplementary materials that help us to reconstruct the dataset. The dataset contain a total of 2,289 buggy functions and each buggy one have its verison of fixed function, so the total is 4578. It has a variety of buggy code types, which include variable misuse, swapped arguments, and incorrect binary operator detection. We randomly split the dataset into training, validation, and testing sets with ratios of 80\%, 10\%, and 10\% respectively.

\subsection{Models}
\textbf{Fine-tuning approach:} We chose two well-known pretrained models for code, which are CodeBERT and CodeT5. We fine-tune the models through several experiments, using various permutations of hyper-parameters including: batch size \{16, 32, 64\} and learning rate \{3-e6, 1-e5, 2-e5, 3-e5\}. We fine-tune the models using the training set, save checkpoints with the lowest validation loss, and then test the models on the testing set. 

\begin{itemize}
    \item \textbf{CodeBERT:} A pretrained model based on a Transformer encoder and follows the same architecture as BERT. This model was pretrained on both source code and natural language. due to the limited resource, we fine-tune codebert-base \footnote{https://huggingface.co/microsoft/codebert-base} with 125 millions of parameters. 
    \item \textbf{CodeT5:} This model, proposed by \citep{codet5}, builds on the T5 (Text-to-Text Transfer Transformer) architecture. CodeT5 was pretrained on the CodeSearchNet data and includes a large dataset of C/C\# programs that were collected from real-world repositories on GitHub.
\end{itemize}

\textbf{In-context learning:}  We consider two language models, GPT-3.5 and CodeLlam, in evaluating our approach. In the in-context Learning approach, selecting demonstration examples is significantly important, so we followed \citep{Jiachang} as he demonstrated an excellent technique for choosing the demonstration examples. We embed all examples from both the training and testing sets using the OpenAI “text-embedding-ada-002” model, which is an exceptionally powerful tool for embedding text and code. Subsequently, we train FAISS using the training set and use the testing set to query and select the nearest examples from the training set based on Euclidean distance.

\begin{itemize}
    \item \textbf{GPT-3.5:} This is one of the most powerful models from OpenAI. We conducted our experiments using "GPT-3.5-turbo," which is one of OpenAI's models boasting a total of 154 billion parameters. It can handle an exceptionally long context of up to 16,385 tokens.
    \item \textbf{CodeLlama:} A large language model for code based on Llama 2. There are two foundational models: CodeLlama-Python, which specializes only in Python, and CodeLlama-Instruct, which is an instruction-following model. All the models are trained on sequences of 16k tokens with 7B, 13B, and 34B parameters each. We use CodeLlama-Instruct with 34B parameters to evaluate our approach.
\end{itemize}

\begin{lstlisting}[style=mystyle,escapechar=!,language=Python,label={real-world_code}, caption={Example of a variable misuse bug found in real-world code.}]
# Buggy code
def __rel_change(self, new: float) -> float:
	if self._likelihoods:
		old = self._likelihoods[-1]
		return abs((new - old) / !\colorbox{light-gray}{old}!)
	return inf
	
# Fixed code
def __rel_change(self, new: float) -> float:
	if self._likelihoods:
		old = self._likelihoods[-1]
		return abs((new - old) / !\colorbox{light-gray}{new}!)
	return inf
\end{lstlisting}

\label{sec:results}
\begin{table*}[h]
\centering
\caption{We evaluated the code-pair classification task for bug detection using a real-world dataset. Our results were compared with those of two baseline methods: the fine-tuning approach and in-context binary classification.}
\label{table:fine-tuning_results}
\resizebox{0.70\textwidth}{!}{%
    \begin{tabular}{l|l|l|cc}
      \toprule
       \textbf{Approach} & \textbf{Tasks}  & \textbf{Models} & \textbf{Accuracy} & \textbf{F1 Score}  \\
      \midrule
      Supervised learning  & Binary  & CodeBERT & 51.96 & 36.41 \\
      (Directly fine-tune) & classification& CodeT5 & 50.00 & 49.67 \\
      \midrule
      Supervised learning  & Binary  & CodeBERT & 61.13 & 60.26 \\
      \citep{alrashedy}& classification& CodeT5 & 60.48 & 59.68 \\
      \midrule
      In-context learning & Binary  & GPT-3.5 & 54.15 & 60.67 \\
      & classification & CodeLlama & 50.44 & 32.24 \\
      \midrule
      In-context learning  & Code-pair  & GPT-3.5 & 72.93 & 84.34 \\
      (Ours) & classification & CodeLlama & 69.87 & 82.26 \\
      \bottomrule
    \end{tabular}%
}
\end{table*}

\section{Experimental Results}
\subsection{Main Results}
\textbf{Fine-tuning results:} We adjust CodeBERT and CodeT5 using the training set by varying hyperparameters such as batch size, learning rate, and number of epochs. Subsequently, The model that had the lowest validation loss was evaluated on the test set. Table \ref{table:fine-tuning_results} presents the results of the binary classification task for both CodeBERT and CodeT5. The accuracy is comparable to random guessing at approximately 50\%, and both models exhibit significantly poor performance on the F1-score. The models are fine-tuned on a small dataset, which makes it difficult for the model to learn the downstream task.

Secondly, there is the multi-stage fine-tuning. First, the models are fine-tuned on a large synthetic dataset for bug detection to learn the domain-specific task. Then, they are further fine-tuned on the PyPIBugs dataset. Overall, this approach shows a 10\% improvement in accuracy performance. It also significantly improves the F1-score, raising it from 36.41 to 60.26 for codeBERT and from 49.67 to 59.68 for CodeT5.

\textbf{In-context (binary classification) results:} To select demonstration examples, we retrieve relevant samples from the training set using FAISS. For each function, we obtain the nearest functions along with their pairs and labels for context. We then input the test function into the model to predict whether the function contains a bug. This task is a binary classification, similar to the previous one, but now we use GPT-3.5 and CodeLlama. GPT-3.5 achieves slightly better performance than a random guess, with an accuracy of 54\%, and the F1-score is around 60\%, comparable to multi-stage fine-tuning. On the other hand, CodeLlama demonstrates poor performance in both accuracy and F1-score.

\textbf{In-context (code-pair classification) results:} Since in-context learning is very sensitive to the demonstration examples, we also retrieve relevant pair examples and prompt the model with two paired functions: the buggy version and the fixed version. We then instruct the model to select the buggy one. The results show a significant improvement in accuracy compared to binary classification. For GPT-3.5, the accuracy increased from 54.15\% to 72.93\%. The accuracy of CodeLlama made an impressive jump from random guessing at 50\% to 69.87\%. The F1 scores for both models are very impressive, standing at 84.34\% and 82.26\% respectively.

\subsection{Error Analysis}
We conducted an experiment on error analysis and found that the model achieves an accuracy of up to 80\% on small functions with fewer than 250 tokens. The model learns and performs better with smaller demonstration examples and inputs. We randomly selected 50 misclassified examples and observed that they contained bugs, specifically of the wrong operator type. We noted that the models struggle to distinguish between buggy functions and their fixed versions when the functions exceed 2000 tokens in length.

In the binary classification task for in-context learning, we ran the experiment three times. The accuracy for GPT-3.5 consistently ranged from 53\% to 56\%, suggesting that the prediction is akin to random guessing. For CodeLlama, the accuracy was 50\%, accompanied by a significant drop in the F1-score.

\section{ Conclusion and Future Work}
We introduced the concept of code-pair classification, a novel approach to bug detection in which Large Language Models (LLMs) are given two versions of a function: one with a bug and the other fixed version. The task for the LLMs is to identify the version containing the bug. This approach was evaluated using two advanced LLMs, GPT-3.5 and CodeLlama. The findings suggest that an LLM is often capable of distinguishing the buggy version from the bug-free one. Furthermore, the task of code-pair classification is much easier compared to being given a snippet and deciding if and where a bug exists.

\section{Limitations}
Our approach assumes that the input to the model consists of a pair of functions: the buggy function and its corrected version. This makes it a much easier task for the model to distinguish the buggy function from the fixed one. For future work, it would be powerful to train the model on pairs of functions rather than on single functions to boost performance. Examples of this include contrastive learning \citep{li2023rethinking} and consider other loss functions such as triplet and hinge losses. 
Secondly, our results in bug detection are still not stellar. This is because the performance of LLM on real-world data tends to be low, as cited in \citep{BUGLAB, great, PLUR}. However, our approach demonstrates an improvement in such situations.

\section*{Acknowledgements}

\bibliography{acl_latex}

\begin{thebibliography}{24}
\expandafter\ifx\csname natexlab\endcsname\relax\def\natexlab#1{#1}\fi

\bibitem[{Ahmad et~al.(2021{\natexlab{a}})Ahmad, Chakraborty, Ray, and
  Chang}]{ahmad-etal-2021-unified}
Wasi Ahmad, Saikat Chakraborty, Baishakhi Ray, and Kai-Wei Chang.
  2021{\natexlab{a}}.
\newblock \href {https://www.aclweb.org/anthology/2021.naacl-main.211} {Unified
  pre-training for program understanding and generation}.
\newblock In \emph{Proceedings of the 2021 Conference of the North American
  Chapter of the Association for Computational Linguistics: Human Language
  Technologies}, pages 2655--2668, Online. Association for Computational
  Linguistics.

\bibitem[{Ahmad et~al.(2021{\natexlab{b}})Ahmad, Chakraborty, Ray, and
  Chang}]{PLBART}
Wasi~Uddin Ahmad, S.~Chakraborty, B.~Ray, and K.~Chang. 2021{\natexlab{b}}.
\newblock Unified pre-training for program understanding and generation.

\bibitem[{Ahmed and Devanbu(2023)}]{Self-Consistency}
Toufique Ahmed and Premkumar Devanbu. 2023.
\newblock Better patching using llm prompting, via self-consistency.

\bibitem[{Allamanis et~al.(2022)Allamanis, H.Jackson-Flux, and
  Brockschmidt}]{BUGLAB}
M.~Allamanis, H.Jackson-Flux, and M.~Brockschmidt. 2022.
\newblock Self-supervised bug detection and repair.

\bibitem[{Alrashedy et~al.(2023)Alrashedy, Hellendoorn, and Orso}]{alrashedy}
Kamel Alrashedy, Vincent~J. Hellendoorn, and Alessandro Orso. 2023.
\newblock Learning defect prediction from unrealistic data.
\newblock \emph{arXiv preprint arXiv:2311.00931}.

\bibitem[{Brown et~al.(2020)Brown, Mann, Ryder, Subbiah, Kaplan, Dhariwal,
  Neelakantan, Sastry, Askell, Agarwa, l~Ariel Herbert-Voss, Krueger, Henighan,
  Child, Ramesh, Ziegler, Jeffrey~Wu, Hesse, Chen, Sigler, Litwin, Gray, Chess,
  Clark, Berner, McCandlish, Radford, Sutskever, and Amodei}]{GPT3}
Tom~B. Brown, Benjamin Mann, Nick Ryder, Melanie Subbiah, Jared Kaplan,
  Prafulla Dhariwal, Arvind Neelakantan, Pranav Shyam~Girish Sastry, Amanda
  Askell, Sandhini Agarwa, l~Ariel Herbert-Voss, Gretchen Krueger, Tom
  Henighan, Rewon Child, Aditya Ramesh, Daniel~M. Ziegler, Clemens~Winter
  Jeffrey~Wu, Christopher Hesse, Mark Chen, Eric Sigler, Mateusz Litwin, Scott
  Gray, Benjamin Chess, Jack Clark, Christopher Berner, Sam McCandlish, Alec
  Radford, Ilya Sutskever, and Dario Amodei. 2020.
\newblock \href {https://doi.org/877-1901.} {Language models are few-shot
  learners}.

\bibitem[{Chakraborty et~al.(2022)Chakraborty, Krishna, Ding, and Ray}]{Saikat}
Saikat Chakraborty, Rahul Krishna, Yangruibo Ding, and Baishakhi Ray. 2022.
\newblock Deep learning based vulnerability detection: Are we there yet?

\bibitem[{Chen et~al.(2023{\natexlab{a}})Chen, Ma, Wang, and
  Cohen}]{chen2022program}
Wenhu Chen, Xueguang Ma, Xinyi Wang, and William~W. Cohen. 2023{\natexlab{a}}.
\newblock Program of thoughts prompting: Disentangling computation from
  reasoning for numerical reasoning tasks.
\newblock \emph{Transactions on Machine Learning Research}.

\bibitem[{Chen et~al.(2023{\natexlab{b}})Chen, Lin, Schärli, and
  Zhou1}]{self-debug}
Xinyun Chen, Maxwell Lin, Nathanael Schärli, and Denny Zhou1.
  2023{\natexlab{b}}.
\newblock Teaching large language models to self-debug.

\bibitem[{Chen et~al.(2022)Chen, Hellendoorn, Lamblin, Maniatis, Manzagol,
  Tarlow, and Moitra}]{PLUR}
Zimin Chen, Vincent~J Hellendoorn, Pascal Lamblin, Petros Maniatis,
  Pierre-Antoine Manzagol, Daniel Tarlow, and Subhodeep Moitra. 2022.
\newblock Plur: A unifying, graph-based view of program learning,
  understanding, and repair.

\bibitem[{Feng et~al.(2020)Feng, D.Guo, Tang, Duan, Feng, Gong, Shou, Qin, Liu,
  and Jiang}]{Codebert}
Z.~Feng, D.Guo, D.~Tang, N.~Duan, X.~Feng, M.~Gong, L.~Shou, B.~Qin, T.~Liu,
  and D.~Jiang. 2020.
\newblock Codebert: A pretrained model for programming and natural languages.

\bibitem[{Fu et~al.(2022)Fu, Tantithamthavorn, Le, Nguyen, and Phung}]{Michael}
Michael Fu, Chakkrit Tantithamthavorn, Trung Le, Van Nguyen, and Dinh Phung.
  2022.
\newblock Vulrepair: A t5-based automated software vulnerability repair.
\newblock In \emph{Proceedings of joint meeting on european software
  engineering conference and symposium on the foundations of software
  engineering}.

\bibitem[{Gao et~al.(2023)Gao, Madaan, Zhou, Alon, Liu, Yang, Callan, and
  Neubig}]{Pal}
Luyu Gao, Aman Madaan, Shuyan Zhou, Uri Alon, Pengfei Liu, Yiming Yang, Jamie
  Callan, and Graham Neubig. 2023.
\newblock Pal: Program-aided language models.

\bibitem[{Hellendoorn et~al.(2020)Hellendoorn, Sutton, Singh, Maniatis, and
  Bieber}]{great}
Vincent~J. Hellendoorn, Charles Sutton, Rishabh Singh, Petros Maniatis, and
  David Bieber. 2020.
\newblock Global relational models of source code.

\bibitem[{Li et~al.(2023)Li, Zhou, Xin, Tuan, Anh, Miao, and
  Chunyan}]{li2023rethinking}
Haochen Li, Zhou, Xin, Tuan, Luu Anh, Miao, and Chunyan. 2023.
\newblock Rethinking negative pairs in code search.
\newblock \emph{arXiv preprint arXiv:2310.08069}.

\bibitem[{Liu et~al.(2023)Liu, Shen, Zhang, Dolan, Carin, and Chen}]{Jiachang}
Jiachang Liu, Dinghan Shen, Yizhe Zhang, Bill Dolan, Lawrence Carin, and Weizhu
  Chen. 2023.
\newblock \href {http://arxiv.org/abs/2101.06804} {What makes good in-context
  examples for gpt-3?}
\newblock arXiv:2101.06804.
\newblock Version 1.

\bibitem[{Madaan et~al.(2023{\natexlab{a}})Madaan, Shypula, Alon, Hashemi,
  Ranganathan, Yang, Neubig, and Yazdanbakhsh}]{madaan2023learning}
Aman Madaan, Alexander Shypula, Uri Alon, Milad Hashemi, Parthasarathy
  Ranganathan, Yiming Yang, Graham Neubig, and Amir Yazdanbakhsh.
  2023{\natexlab{a}}.
\newblock Learning performance-improving code edits.
\newblock \emph{arXiv preprint arXiv:2302.07867}.

\bibitem[{Madaan et~al.(2023{\natexlab{b}})Madaan, Tandon, Gupta, Hallinan,
  Gao, Wiegreffe, Alon, Dziri, Prabhumoye, Yang, Gupta, Majumder, Hermann,
  Welleck, Yazdanbakhsh, and Clark}]{Self-refine}
Aman Madaan, Niket Tandon, Prakhar Gupta, Skyler Hallinan, Luyu Gao, Sarah
  Wiegreffe, Uri Alon, Nouha Dziri, Shrimai Prabhumoye, Yiming Yang, Shashank
  Gupta, Bodhisattwa~Prasad Majumder, Katherine Hermann, Sean Welleck, Amir
  Yazdanbakhsh, and Peter Clark. 2023{\natexlab{b}}.

\bibitem[{Raffel et~al.(2020)Raffel, Shazeer, Roberts, Lee, Narang, Matena,
  Zhou, Li, and Liu}]{2020t5}
Colin Raffel, Noam Shazeer, Adam Roberts, Katherine Lee, Sharan Narang, Michael
  Matena, Yanqi Zhou, Wei Li, and Peter~J. Liu. 2020.
\newblock \href {http://jmlr.org/papers/v21/20-074.html} {Exploring the limits
  of transfer learning with a unified text-to-text transformer}.
\newblock \emph{Journal of Machine Learning Research}, 21(140):1--67.

\bibitem[{Rozière et~al.(2023)Rozière, Gehring, Gloeckle, Sootla, Gat, Tan,
  Adi, Liu, Remez, Rapin, Kozhevnikov, Evtimov, Bitton, Bhatt, Ferrer,
  Grattafiori, Xiong, Défossez, Copet, Azhar, Touvron, Martin, Usunier,
  Scialom, and Synnaeve}]{Code_llama}
Baptiste Rozière, Jonas Gehring, Fabian Gloeckle, Sten Sootla, Itai Gat,
  Xiaoqing~Ellen Tan, Yossi Adi, Jingyu Liu, Tal Remez, Jérémy Rapin, Artyom
  Kozhevnikov, Ivan Evtimov, Joanna Bitton, Manish Bhatt, Cristian~Canton
  Ferrer, Aaron Grattafiori, Wenhan Xiong, Alexandre Défossez, Jade Copet,
  Faisal Azhar, Hugo Touvron, Louis Martin, Nicolas Usunier, Thomas Scialom,
  and Gabriel Synnaeve. 2023.
\newblock Code llama: Open foundation models for code.

\bibitem[{Vaswani et~al.(2017)Vaswani, Shazeer, Parmar, Uszkoreit, Jones,
  Gomez, Łukasz Kaiser, and Polosukhin}]{Vaswani_17}
Ashish Vaswani, Noam Shazeer, Niki Parmar, Jakob Uszkoreit, Llion Jones,
  Aidan~N Gomez, Łukasz Kaiser, and Illia Polosukhin. 2017.
\newblock Attention is all you need.
\newblock In \emph{Advances in neural information processing systems}.

\bibitem[{Wang et~al.(2024)Wang, Peng, Gao, Chen, Wang, Gao, and
  Lyu}]{Multi-Intent}
Chaozheng Zongjie~Li Wang, Yun Peng, Shuzheng Gao, Sirong Chen, Shuai Wang,
  Cuiyun Gao, and Michael~R. Lyu. 2024.
\newblock Large language models are few-shot summarizers: Multi-intent comment
  generation via in-context learning.

\bibitem[{Wang et~al.(2021)Wang, Wang, Joty, and Hoi}]{codet5}
Yue Wang, W.~Wang, S.~Joty, and Steven~CH Hoi. 2021.
\newblock Codet5: Identifier-aware unified pre-trained encoder-decoder models
  for code understanding and generation.

\bibitem[{Zhou et~al.(2023)Zhou, Alon, Xu, Wang, Jiang, and
  Neubig}]{zhou23docprompting}
Shuyan Zhou, Uri Alon, Frank~F. Xu, Zhiruo Wang, Zhengbao Jiang, and Graham
  Neubig. 2023.
\newblock \href {https://arxiv.org/abs/2207.05987} {Docprompting: Generating
  code by retrieving the docs}.
\newblock In \emph{International Conference on Learning Representations
  (ICLR)}, Kigali, Rwanda.

\end{thebibliography}

\appendix

\onecolumn


\end{document}